\documentclass[reprint, superscriptaddress, amsmath, amssymb, aps, pra]{revtex4-2}
\usepackage{graphicx}
\usepackage{dcolumn} 
\usepackage{xcolor}
\usepackage[absolute]{textpos}
\usepackage{bm}
\usepackage{bbm}
\usepackage{braket}
\definecolor{darkblue}{rgb}{0,0.3,0.7}
\usepackage{hyperref, comment}
\usepackage{float}
\hypersetup{
colorlinks=true,
linkcolor=darkblue,
filecolor=blue,
citecolor=darkblue,  
urlcolor=darkblue,
}

\DeclareMathOperator{\Tr}{Tr}
\def\abar{{\overline{a}}}       
\def\bbar{{\overline{b}}}
\def\bm{\mathbf{m}}
\def\bn{\mathbf{n}}
\def\bp{\mathbf{p}}
\def\cV{\mathcal{V}}
\def\cF{\mathcal{F}}

\begin{document}

\preprint{APS/123-QED}

\title{Developing a practical model for noise in entangled photon detection}
\author{Taman Truong}
\email{tvtruon1@asu.edu}
\author{Christian Arenz}
\affiliation{School of Electrical, Computer, and Energy Engineering and Research Technology Office, Arizona State University, Tempe, Arizona 85287, USA}
\author{Joseph~M. Lukens}
\email{jlukens@purdue.edu}
\affiliation{School of Electrical, Computer, and Energy Engineering and Research Technology Office, Arizona State University, Tempe, Arizona 85287, USA}
\affiliation{Elmore Family School of Electrical and Computer Engineering and Purdue Quantum Science and Engineering Institute, Purdue University, West Lafayette, Indiana 47907, USA}
\affiliation{Quantum Information Science Section, Oak Ridge National Laboratory, Oak Ridge, Tennessee 37831, USA}

\date{\today}

\begin{abstract}
We develop a comprehensive model for the effective two-photon density matrix produced by a parametric source of entangled photon pairs under 
a variety of detector configurations commonly seen in a laboratory setting: two and four photon number-resolving (PNR) and threshold detectors. We derive the probability of obtaining a single coincidence assuming 
Poisson-distributed photon pairs, non-unit detection efficiency, and dark counts; obtain the effective density matrix; and use this quantity to compute the fidelity of this state 
with respect to the maximally entangled ideal. The 4 PNR case admits an analytic result valid for any combination of parameters, while all other cases leverage low-efficiency approximations to arrive at closed-form expressions. 
Interestingly, our model reveals appreciable fidelity improvements from four detectors as opposed to two 
yet minimal advantages for PNR over threshold detectors in the regimes explored. Overall, our work provides a valuable tool for the quantitative design of two-photon experiments under realistic nonidealities.
\end{abstract}

\maketitle

\section{Introduction}
\label{sec:intro}
Analyzing photon statistics is a ubiquitous process in quantum optics, as photon detection is central to characterizing, understanding, and optimizing quantum light sources. 
Historically, threshold detectors (which can only distinguish between vacuum and $\geq 1$ photon) have dominated the field~\cite{hadfield2009single, Eisaman2011}, 
although the need for true PNR detectors has become increasingly acute as photonic quantum information has progressed, representing key components in applications such as linear optical quantum computing~\cite{Knill2001, Kok2007}, Gaussian boson sampling (GBS)~\cite{Hamilton2017, Kruse2019}, and the heralded production of non-Gaussian resource states 
such as Gottesman--Kitaev--Preskill qubits~\cite{Gottesman2001,Eaton2019,Tzitrin2020,Takase2023}.
Technologically speaking, transition-edge sensors (TESs) have led the way in PNR detection, supporting intrinsic photon-number discrimination up to dozens of photons~\cite{Rosenberg2005, Harder2016,Arrazola2021,Madsen2022}. More recently, superconducting nanowire single-photon detectors (SNSPDs)---for many years the leader in threshold photon detection~\cite{Natarajan_2012, FerrariSchuckPernice+2018+1725+1758, You+2020+2673+2692, Korzh2020, Reddy:20, Chang2021, 10.1063/5.0045990}---have also emerged as valuable PNR detectors, with parallel arrays now being explored in a variety of photonic quantum information processing contexts~\cite{Cheng2022, Alexander2024, Stasi2025} 
In any case, both PNR and threshold detectors face nonidealities, whether internal to the devices themselves (like imperfect detection efficiency and dark counts) or external (e.g., channel losses and background light), that significantly impact the ability for these devices to probe quantum states accurately and efficiently.

For experiments in which the probability of detection within a resolving time is low, the impact of accidental coincidences can be well modeled by the ``product-of-singles'' formula \cite{PhysRev.53.752, 10.1119/1.3354986}, which states that the rate of simultaneous random clicks on two detectors is proportional to the product of the individual rates on each. Both intuitive and highly accurate under many typical experimental conditions, this rule has proven itself a workhorse in quantum optics. Yet, in many cases of interest, it is possible to 
derive an even more informative summary of the noise through an effective density matrix: ``effective'' in the sense that it can account for all outcomes of an experiment in a simplified Hilbert space. For example, in modeling the detection of two-photon entanglement from spontaneous parametric down-conversion (SPDC) 
\cite{Mandel1995, rubin1996transverse, keller1997theory, Shih2003, Couteau2018, schneeloch2019introduction, zhang2021spontaneous},
the complete Hilbert space including multipair emission, multiple electromagnetic modes, and spurious detector clicks can frequently be reduced to a density matrix in an effective two-qubit Hilbert space. Although such Hilbert space compression is not always possible, when it is, a potentially complex problem can be reduced to a simple density matrix that reflects all impairments in the system.

Critical in this regard, Takesue and Shimizu~\cite{Takesue2009EffectsOM} have developed useful formulas describing the effective state 
of indistinguishable and distinguishable entangled photon pairs generated by parametric processes such as SPDC, assuming the use of two imperfect threshold detectors. 
Considering a representative two-photon interference setup, equations for the coincidence rates, interferometric visibilities, and the resulting density matrices at the high and low fringes are derived in terms of the average pair number and detection efficiency, and then expanded to include the effects of dark counts after approximations of low detection efficiency have been applied. 
The paper does not address PNR detectors as an option to analyze entangled photon detection, nor does it compare effects from the quantity of detectors typically considered for two photonic qubits, namely two or four. In light of the growing importance of PNR detection in modern quantum optics, there exists strong motivation to expand the reach of Takesue and Shimizu's highly useful formalism into more general regimes of operation.

In this paper, we develop such an updated model based on basic probability theory to obtain the effective detected quantum state of two entangled photons from SPDC under a variety of experimentally relevant conditions. Focusing on the relative merits of PNR versus threshold detectors and two-detector versus four-detector setups, we derive the probability of obtaining a single coincidence given an arbitrary number of photons, utilizing this expression to obtain an effective density matrix in terms of the dark count probability, detector efficiency, and average photon pair number under four configurations: four PNR detectors, four threshold detectors, two PNR detectors, and two threshold detectors. 
To maintain a manageable scope and reveal the main points of interest, we focus on distinguishable (i.e., Poisson-distributed) photon pairs and identical channels and detectors for both photons, yet our approach can easily be modified to account for other distributions and asymmetric components. 
Surprisingly, we find an exact solution for the case of four PNR detectors valid for \emph{any} parameter combination. For all other cases, simplification to an effective density matrix requires typical assumptions of low pair rate and low combined channel and detector efficiencies, the accuracy of which we confirm through numerical simulations under common regimes of operation. 

Our results yield the interesting conclusion: while four detectors appreciably improve the postselected two-photon state by filtering out unwanted events that two detectors alone cannot see, PNR detectors provide negligible enhancements in the regimes explored---an ostensibly surprising result given their ability to filter out multiphoton events, yet intuitively reflecting the fact that the dominant multipair noise stems from photon loss in the linearized conditions explored.

The paper is organized as follows. Sec.~\ref{sec:Prelim} delineates the problem statement and model, while 
Sec.~\ref{sec:results} derives the effective density matrices for each case. 
In Sec.~\ref{sec:analysisresults}, we validate the suitability of the approximations taken in Sec.~\ref{sec:results} through visibility comparisons and then analyze the behavior of the effective quantum states as parameters are tuned.
Sec.~\ref{sec:conclusion} summarizes the results and explores potential areas of improvement within our mathematical model for entanglement using photon detection systems. 
\section{Preliminaries} \label{sec:Prelim}
\subsection{Problem Formulation}
\label{sec:probForm}
\begin{figure}[!tb]
  \centering
  \includegraphics[width=1.0\columnwidth]{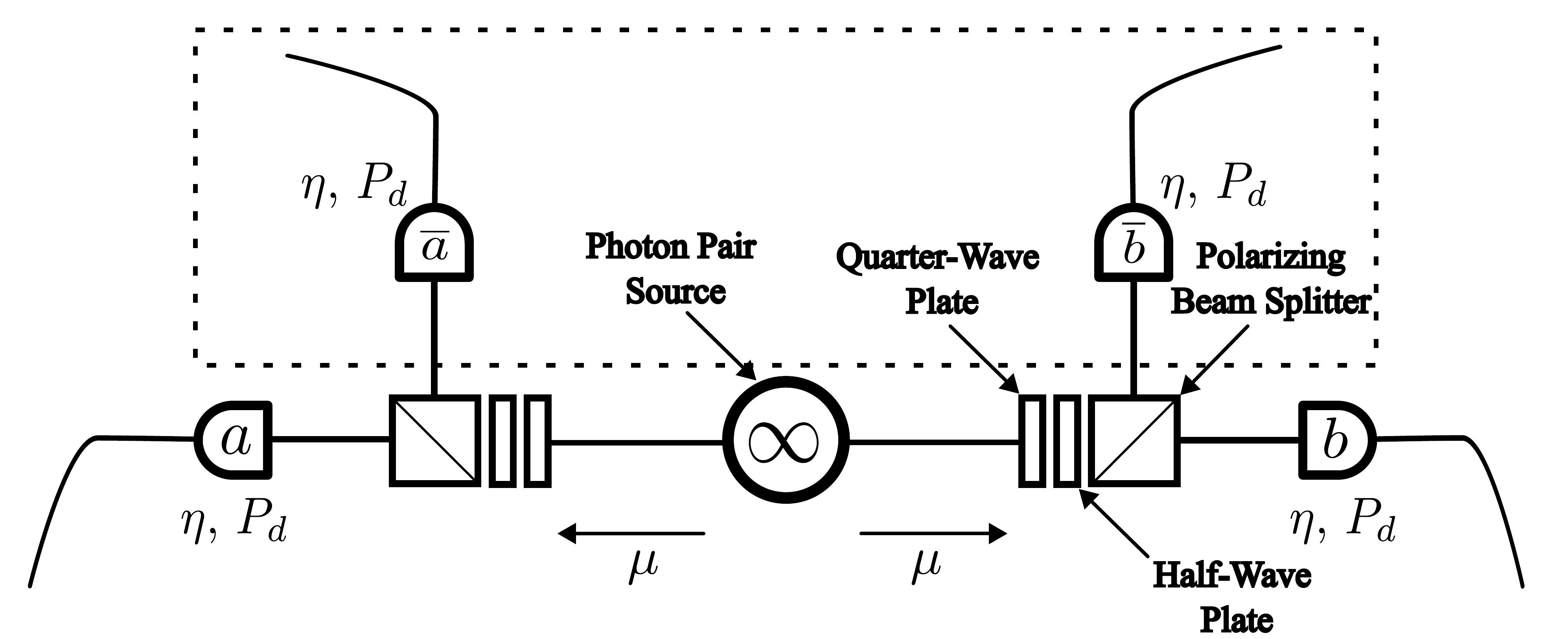}
  \caption{Envisioned setup for two-qubit entangled-photon detection in basis $\{\ket{a},\ket{\abar}\}$ for one photon,  $\{\ket{b},\ket{\bbar}\}$ for the other. For a four-detector setup, all shown detectors are used, whereas a two-detector setup excludes those in the dotted rectangular region. We consider a pair generation probability per timeslot of $\mu$ and identical efficiency $\eta$ (channel and detector) and dark count probability $P_d$ for all lightpaths.}
  \label{fig:1}
\end{figure}

Consider an entangled photon source that we wish to measure in a specific pair of bases  as shown in Fig.~\ref{fig:1}. In each timeslot (defined by, e.g., the pump pulse or the system resolving time), the central source produces $x$ photon pairs from a Poisson distribution with mean $\mu$. This model assumes each pair is in principle distinguishable (i.e., populates a distinct time-frequency mode), yet the detection system cannot resolve them---typical for narrowband-pumped SPDC with THz-scale marginal photon bandwidths.
The signal modes (moving left in Fig.~\ref{fig:1}) are measured in some qubit basis $\{\ket{a},\ket{\overline{a}}\}$, while the idler modes (traveling right in Fig.~\ref{fig:1}) are measured in $\{\ket{b},|\overline{b}\rangle\}$. 
We define $\eta\in[0,1]$ as the probability for a given photon in the respective state to be detected (incorporating both channel and detection efficiency), while $P_d\in[0,1]$ sets the probability of measuring a dark count.

Both bases are visually represented as polarization states, which can be measured in an arbitrary pair of bases using only wave plates and polarizing beam splitters. Yet the theory we develop applies to any dual-rail encoding---e.g., path, time bins, frequency bins, or orbital angular momentum---where qubits are represented as single photons occupying a superposition two encoding modes. These encoding modes can be further classified into time-frequency modes that are distinct yet unresolvable by the detection system. Thus, measurement in the qubit basis $\{\ket{a},\ket{\abar}\}$ should be formally understood as the incoherent sum of all spectro-temporal modes sharing the encoding mode $\ket{a}$ or $\ket{\abar}$, the precise meaning of which will be defined in the following sections.

For a given timeslot, we define $\bm=(m_{ab}, m_{a\bbar}, m_{\abar b}, m_{\abar\bbar})$ as the ground truth number of photon pairs projected onto each joint setting (in the absence of loss), which accordingly satisfies $m_{ab} + m_{a\bbar} + m_{\abar b} + m_{\abar\bbar} = x$. We then seek to find the probability $c_i(\bm)$ of a coincidence between states $\ket{a}$ and $\ket{b}$ of the radiation
field in the target Hilbert space for each of the following four cases $i\in\{1,2,3,4\}$ that specify the types and number of detectors used:
\begin{itemize}
    \item[(1)]
    For four PNR detectors: 
    \begin{equation}
    \label{eq:case1}
        c_1(\bm) = \Pr(\bn|\bm), \text{ where } \bn = (1, 0, 1, 0).
    \end{equation}
    \item[(2)]
    For four threshold detectors:
    \begin{equation}
    \label{eq:case2}
        c_2(\bm) = \sum_{n_a = 1}^\infty\sum_{n_b = 1}^\infty \Pr(\bn | \bm), \text{ where } \bn = (n_a, 0, n_b, 0).
    \end{equation}
    \item[(3)]
    For two PNR detectors:
    \begin{equation}
    \label{eq:case3}
        c_3(\bm) = \sum_{n_\abar = 0}^\infty\sum_{n_\bbar = 0}^\infty \Pr(\bn|\bm), \text{ where } \bn = (1, n_{\abar}, 1, n_\bbar).
    \end{equation}
    \item[(4)]
    For two threshold detectors:
    \begin{multline}
    \label{eq:case4}
        c_4(\bm) = \sum_{n_a = 1}^\infty\sum_{n_\abar = 0}^\infty\sum_{n_b = 1}^\infty\sum_{n_\bbar = 0}^\infty \Pr(\bn|\bm), \\ \text{ where } \bn = (n_a, n_\abar, n_b, n_\bbar).
    \end{multline}
\end{itemize}
We use $\bn = (n_a, n_\bbar, n_b, n_\bbar)$ to denote the number of clicks observed by the four PNR detectors. Thus, a threshold detector can be viewed as the special case of summing over $n_i\geq 1$, while the absence of a detector follows by summing over all outcomes ($n_i\geq 0$)~\cite{Kok2007}. This identification allows all four cases to rely on the same basic probability $\Pr(\bn|\bm)$.

Our definition of a coincidence is precisely two clicks: one at $\ket{a}$ and the other at $\ket{b}$. In the four-detector scenarios, this requires that no clicks be found for the states $\ket{\overline{a}}$ and $|\overline{b}\rangle$ as well. The total coincidence probability $C_i$ for each case $i$ follows by summing over all possible $\bm$ and photon pairs $x$ as
\begin{equation}
\label{eq:totalCoinc}
C_i = \sum_{x=0}^\infty  \Pr(x) \sum_{\bm(x)} c_i (\bm) \Pr(\bm|x),
\end{equation}
where $\bm(x)$ denotes all $\bm\in\mathbb{N}_0^4$ such that $m_{ab}+m_{a\bbar}+m_{\abar b} + m_{\abar\bbar} = x$.
From this probability, we seek to find the effective density matrix $\rho_i$ such that 
\begin{equation}
\label{eq:rhoEff}
C_i \propto \bra{ab}\rho_i\ket{ab},
\end{equation}
where $\ket{ab} = \ket{a} \otimes \ket{b}$. Accordingly, the rest of this paper can be summarized as solving and analyzing Eqs.~(\ref{eq:case1}--\ref{eq:rhoEff}) for a specific triad of probability mass functions (PMFs): $\Pr(\bn|\bm)$, $\Pr(\bm|x)$, and $\Pr(x)$.

It is important to note that $C_i$ in Eq.~(\ref{eq:totalCoinc}) is always well defined and can be computed for any combination of parameters, yet there is no guarantee that it can be written as $C_i \propto\bra{ab}\rho_i\ket{ab}$ as needed in Eq.~(\ref{eq:rhoEff}) to define an effective density matrix. The \emph{physical} Hilbert space consists of many photon pairs and time-frequency modes, whereas the \emph{effective} Hilbert space for $\rho_i$ considers just two qubits. 
Accordingly, $C_i$ need not be linear in the two-qubit measurement operator $\ket{ab}\bra{ab}$ in the larger Hilbert space. Nonetheless, as we will see in the following sections, $\rho_i$ can be exactly defined for the four PNR case, and derived under reasonable parameter approximations in the other three.

Incidentally, our objective of deriving an effective low-dimensional representation of a higher-dimensional experiment is highly related to squashing models, which comprise linear maps that preserve measurement statistics for a specified set of observables when converting from a higher- to a lower-dimensional Hilbert space~\cite{Beaudry2008,Moroder2010,Fung2011,Gittsovich2014}. Squashing's criterion of statistical equivalence is precisely that of our desired linear formulation $C_i \propto\bra{ab}\rho_i\ket{ab}$, namely, that a lower-dimensional density matrix $\rho_i$ can reproduce the observations from the complete physical model. Nevertheless, because squashing operations have historically been developed for adversarial, or at least skeptical, settings---such as quantum key distribution~\cite{Beaudry2008, Fung2011, Gittsovich2014} and entanglement verification~\cite{Moroder2010}---they demand validity for \emph{arbitrary} inputs on the Hilbert space. In contrast, our effort focuses on the differences in multiple measurement schemes for a \emph{specific} optical state, unlocking more flexibility in the theoretical development.

To elaborate on this point, we note that while Eqs.~(\ref{eq:case1}--\ref{eq:totalCoinc}) hold in general for any source of photon pairs, the development in this paper concentrates on a particular ground truth quantum state, which can be written in the full Hilbert space as the tensor product of contributions in $T$ time-frequency modes:
\begin{equation}
\label{eq:tensorProd}
\rho_{\mathrm{full}} = \bigotimes_{t=1}^T \left[\left(1-\frac{\mu}{T}\right)\ket{\mathrm{vac}}\bra{\mathrm{vac}} + \frac{\mu}{T}\rho_{AB} \right]_t, 
\end{equation}
where each index $t$ corresponds to a specific quadruple of spatio-spectral modes $(a_t,\abar_t, b_t,\bbar_t)$, $\ket{\mathrm{vac}}$ denotes the vacuum state in said modes, and $\rho_{AB}$ signifies the quantum state of two photons generated in those modes.

The state in Eq.~(\ref{eq:tensorProd}) assumes a low probability of pair generation per time-frequency mode ($\mu/T$) and identical states within each---reasonable for the regime $\mu\ll 1$ and photons are generated by a narrowband pump (e.g., $T\approx 10^6$ for a 1~MHz linewidth laser and 1~THz phase-matching bandwidth~\cite{Chekhova2011}). A Poisson distribution in the number of pairs $x$ [i.e., the number of $\rho_{AB}$ factors in an expansion of Eq.~(\ref{eq:tensorProd})] then follows in the limit $T\rightarrow \infty$~(cf. Sec. 13.3.2 of Ref.~\cite{Mandel1995} and Sec. 2.2 of Ref.~\cite{Takesue2009EffectsOM}). Our approach can therefore be viewed as formally replacing a $5T$-dimensional ground truth $\rho_{\mathrm{full}}$ (vacuum plus four dimensions per time-frequency mode) by a four-dimensional equivalent $\rho_i$ that preserves all measurement statistics in the two-qubit space of interest. Yet instead of adopting the language of squashing maps and positive operator-valued measures to perform this reduction, the specialization to Eq.~(\ref{eq:tensorProd}) allows us to proceed directly with conditional probabilities, which we find  more intuitive and straightforward.

\subsection{Probability Mass Functions (PMFs)}
\label{sec:PMFs}
Starting with $\Pr(\bn|\bm)$, we first note that, conditioned on the ground truth $\bm$, events at each detector are independent. Hence we can break up the joint detection probability  as
\begin{equation}
\label{eq:marginals}
    \Pr(\bn|\bm) = \Pr(n_a| m_a)\Pr(n_\abar | m_\abar)\Pr(n_b|m_b)\Pr(n_\bbar|m_\bbar),
\end{equation}
where for convenience we have defined the total number of single photons destined for each detector as
\begin{align}
\label{eq:singlePhotons}
    m_a &= m_{ab} + m_{a\bbar}, &  m_\abar &= m_{\abar b} + m_{\abar\bbar}, \nonumber\\
    m_b &= m_{ab} + m_{\abar b}, & m_\bbar &= m_{a\bbar} + m_{\abar\bbar},
\end{align}
which must satisfy $m_a + m_{\abar} = m_b + m_{\bbar} = x$; i.e., all individual photons generated must project onto either of the two outcomes for each qubit basis.

To obtain $\Pr(n_i|m_i)$, we must select an appropriate dark count model. Although both Poisson~\cite{Lee2004} and thermal~\cite{Scherer2009} distributions have been explored in this context, we enlist a particularly intuitive, recently proposed Bernoulli model in which at most one dark count can be generated per timeslot~\cite{PhysRevApplied.17.034071}. Combining this with the binomial distribution associated with detecting the photons themselves~\cite{Lamb1969}, the probability of experimentally detecting $n_i$ clicks at detector $i$ given $m_i$ incident photons becomes
\begin{multline}
\label{eq:Bernoulli}
    \Pr(n_i|m_i) = P_d\binom{m_i}{n_i - 1} \eta^{n_i - 1}(1 - \eta)^{m_i - (n_i - 1)} \\
    + (1 - P_d)\binom{m_i}{n_i} \eta^{n_i}(1 - \eta)^{m_i - n_i},
\end{multline}
where we adopt the convention $\binom{n}{k}=0$ for $k>n$.
In words, $n_i$ clicks can result either from $n_i-1$ photons and one dark count (first term) or from $n_i$ photons and no dark counts (second term). With this result, we see immediately from $\displaystyle\sum_{n_i = 0}^\infty \Pr(n_i|m_i) = 1$ that
\begin{equation}
    \sum_{n_i = 1}^\infty \Pr(n_i \mid m_i) = 1 - (1 - P_d)(1 - \eta)^{m_i}
\end{equation}
returns the threshold detector probability as the complement of not receiving any clicks.

As an aside, an arbitrary dark count distribution $D(k)$ ($k\in\mathbbm{N}_0$) could be incorporated into $\Pr(n_i|m_i)$ as~\cite{Lee2004}
\begin{equation}
\label{eq:darkCount}
\Pr(n_i|m_i) = \sum_{k=0}^{n_i} D(k) \binom{m_i}{n_i-k} \eta^{n_i-k}(1-\eta)^{m_i-(n_i-k)},
\end{equation}
which again reflects the intuitive understanding of $n_i$ clicks resulting from $k$ dark counts and $n_i-k$ photons. In practice, useful single-photon detectors attain $D(0)\approx 1$, with $D(1)$ a small correction and $D(k)$ vanishingly small for $k\geq 2$. As representative (but nonexhaustive) examples, state-of-the art TESs~\cite{Manenti2024} and SNSPDs~\cite{Hochberg2019} have reached dark count rates $\lesssim$10$^{-4}$~s$^{-1}$; typical commercial SNSPDs routinely output $\lesssim$100~s$^{-1}$ dark counts~\cite{QuantumOpus2025}; and even free-running commercial single-photon avalanche photodiodes show $\lesssim$200~s$^{-1}$ at 10\% quantum efficiency~\cite{IDQuantique2025}. In a relatively large 10~ns window, for example, these numbers correspond to $D(1)\in[10^{-12},10^{-6}]$, making a Bernoulli random variable with $P_d=D(1)$ an excellent approximation for any ground truth distribution (Poisson, thermal, or otherwise). Therefore, while additional analyses with more general $D(k)$ would be valuable directions for future studies, they are expected to have minimal impact on any practical conclusions drawn from our model.

In consequence of the independence of each photon pair for the distinguishable case, the probability of obtaining the ground truth projection $\bm$ given $x$  pairs is described by the multinomial distribution
\begin{equation}
\label{eq:multinomial}
    \Pr(\bm|x) = \frac{x!}{m_{ab}! m_{\abar b}! m_{a\bbar}! m_{\abar\bbar}!}p_{ab}^{m_{ab}} p_{\overline{a}b}^{m_{\overline{a}b}} p_{a\overline{b}}^{m_{a\overline{b}}} p_{\overline{a}\overline{b}}^{m_{\overline{a}\overline{b}}},
\end{equation}
where the probabilities $\bp=(p_{ab}, p_{a\bbar}, p_{\abar b}, p_{\abar\bbar})$ are defined as
\begin{align}
    p_{ab} &= \braket{ab | \rho_{AB} | ab}, & p_{\abar b} &= \braket{\abar b | \rho_{AB}| \abar b}, \nonumber\\
    p_{a\bbar} &= \braket{a\bbar | \rho_{AB} | a\bbar}, & p_{\abar\bbar} &= \braket{\abar\bbar|\rho_{AB}|\abar\bbar}, 
\end{align}
for the relevant ``single-pair'' density matrix $\rho_{AB}$---\emph{not} the effective density matrix of interest $\rho_i$, but rather the two-photon state in one time-frequency mode $t\in\{1,...,T\}$ [Eq.~(\ref{eq:tensorProd})]. (Since $\rho_{AB}$ does not depend on $t$, and the detector cannot resolve it, we suppress it for clarity in all that follows.) We can similarly define marginal probabilities as
\begin{align}
\label{eq:marginalP}
    p_a &= p_{ab} + p_{a\bbar} = \braket{a|\rho_A|a}, & p_\abar &= p_{\abar b} + p_{\abar\bbar} = \braket{\abar|\rho_A|\abar}, \nonumber\\
    p_b &= p_{ab} + p_{\abar b} = \braket{b|\rho_B|b}, & p_\bbar &= p_{a\bbar} + p_{\abar\bbar} = \braket{\bbar|\rho_B|\bbar},
\end{align}
where $\rho_A \equiv \Tr_B \rho_{AB}$ and $\rho_B \equiv \Tr_A \rho_{AB}$ denote the marginal single-photon density matrices. These probabilities define what would be obtained in the ideal case of a single pair with no detector noise.
Finally, due to the assumption of independent, identically distributed photon pairs with the generation probability $\mu/T$ for each time-frequency mode, we obtain a Poisson distribution for the total number of pairs $x$:
\begin{equation}
\label{eq:Poisson}
\Pr(x) = e^{-\mu}\frac{\mu^x}{x!}.
\end{equation}

Before proceeding, it is useful to pause and highlight the simplifications made so far. The overall formalism introduced in Eqs.~(\ref{eq:case1}--\ref{eq:totalCoinc}) makes no assumptions about the photon statistics, quantum channels, or detector characteristics. Yet in moving to Sec.~\ref{sec:PMFs}, three main assumptions are leveraged to select concrete PMFs: (i)~identical channels and detectors $(\eta,P_d)$ with Bernoulli dark counts [Eq.~(\ref{eq:Bernoulli})], (ii)~independent photon pairs [Eq.~(\ref{eq:multinomial})], and (iii)~Poisson-distributed generation [Eq.~(\ref{eq:Poisson})]. Therefore all subsequent results rely on these assumptions, but we emphasize that they can easily be removed by specifying alternative PMFs such as, e.g., thermally distributed photons in the four measured modes of interest~\cite{Takesue2009EffectsOM}, making our formalism adaptable to other typical scenarios.
\section{Results}
\label{sec:results}
In this section, we summarize the coincidence probabilities and effective density matrices obtained from the model and assumptions delineated in Sec.~\ref{sec:Prelim}. Additional details can be found in Appendix~\ref{Appendix}.

\subsection{Case 1 --- 4 PNR Detectors} \label{sec:4PNR}
In the case of 4 PNR detectors, a single coincidence is registered when the $\ket{a}$ and $\ket{b}$ ports detect one click, and $\ket{\abar}$ and $\ket{\bbar}$ detect no click, i.e., $\bn = (1, 0, 1, 0)$.
The probability of detecting a single coincidence given ground truth photon vector $\bm$ is
\begin{widetext}
\begin{equation}
\begin{split}
\label{eq:c1}
    c_1(\bm) & = \Pr(1|m_a)\Pr(0|m_\abar)\Pr(1|m_b)\Pr(0|m_\bbar)\\
     & = (1 - P_d)^2(1 - \eta)^{2(x - 1)}[P_d(1-\eta) + (1 - P_d)\eta m_a][P_d(1-\eta) + (1 - P_d)\eta m_b],
\end{split}
\end{equation}
which after summing over all $\bm$ and $x$ returns the total coincidence probability (see Appendix~\ref{app4PNR})
\begin{equation}
\label{eq:C1v1}
C_1 =  (1 - P_d)^2e^{\mu[(1 - \eta)^2 - 1]} 
\Big\{\underbrace{[P_d + (1 - P_d)\mu\eta(1 - \eta)p_a][P_d + (1 - P_d)\mu\eta(1 - \eta)p_b]}_{\text{accidental coincidences}}
+ \underbrace{(1 - P_d)^2\mu\eta^2 p_{ab}}_{\text{correlated coincidences}}\Big\}.
\end{equation}
No approximations were applied to reach this point, yet the result assumes a simple form, featuring a ``correlated coincidences'' term scaling like the joint probability for a single pair $p_{ab}$ and an ``accidental coincidences'' term comprising all other possibilities. The latter is very similar to the standard product-of-singles expression for the regime $\eta,P_d,\mu\ll 1$, but with $(1-\eta)$ and $(1-P_d)$ correction factors that ensure validity for all $\eta,P_d\in[0,1]$ and $\mu>0$. 

Because of the linearity in probabilities $p_a$, $p_b$, and $p_{ab}$, we can immediately replace them with single-pair density matrices as
\begin{equation}
\label{eq:C1v2}
C_1 = (1 - P_d)^2e^{\mu[(1 - \eta)^2 - 1]}
    \bra{ab}[P_d\mathbbm{1}_A + (1 - P_d)\mu\eta(1 - \eta)\rho_A] \otimes [P_d\mathbbm{1}_B + (1 - P_d)\mu\eta(1 - \eta)\rho_B] + (1 - P_d)^2\mu\eta^2\rho_{AB}\ket{ab},
\end{equation}
where $\mathbbm{1}_A$ ($\mathbbm{1}_B$) denotes the $2\times 2$ identity matrix in the effective Hilbert space of the signal (idler). 

Significantly, because the qubit measurement bases $\{\ket{a},\ket{\abar}\}$ and $\{\ket{b},\ket{\bbar}\}$ are completely arbitrary, this expression  holds for any separable pure state projection $\ket{ab}\equiv \ket{a}_A\otimes\ket{b}_B = (\cos\alpha\ket{0} + e^{i\phi}\sin\alpha\ket{1})_A \otimes (\cos\beta\ket{0} + e^{i\varphi}\sin\beta\ket{1})_B$, where $\ket{0}$ and $\ket{1}$ comprise the computational basis for each qubit. Accordingly, 
we can generalize Eq.~(\ref{eq:C1v2}) to $C_1\propto\braket{ab|\rho_1|ab}$ via the effective density matrix
\begin{equation}
\label{eq:rho1}
\rho_1 = \frac{1}{K_1} \Big\{ [P_d\mathbbm{1}_A + (1 - P_d)\mu\eta(1 - \eta)\rho_A] \otimes [P_d\mathbbm{1}_B + (1 - P_d)\mu\eta(1 - \eta)\rho_B] + (1 - P_d)^2\mu\eta^2\rho_{AB} \Big\},
\end{equation}
where 
\begin{equation}
\label{eq:K1}
K_1 = [2P_d + (1 - P_d)\mu\eta(1 - \eta)]^2 + (1 - P_d)^2\mu\eta^2
\end{equation}
ensures normalization $\Tr\rho_1 =1$. This effective density matrix enjoys the precise meaning of producing results that are statistically equivalent to the full physical model; i.e., it accurately predicts coincidence probabilities for all two-qubit measurements in the setup of Fig.~\ref{fig:1} under the PMFs assumed in Sec.~\ref{sec:PMFs}.

\subsection{Case 2 --- 4 Threshold Detectors}
\label{sec:4thr}
With 4 threshold detectors, a single coincidence is registered when the virtual PNR detectors at $\ket{a}$ and $\ket{b}$ receive at least one click, while $\ket{\abar}$ and $\ket{\bbar}$ report no clicks. Therefore the coincidence probability conditioned on $\bm$ is 
\begin{equation}
\begin{split}
\label{eq:c2}
c_2(\bm) & = \left[ \sum_{n_a = 1}^\infty\Pr(n_a|m_a) \right] \Pr(0|m_\abar) \left[\sum_{n_b = 1}^\infty\Pr(n_b|m_b) \right] \Pr(0|m_\bbar) \\
  &= (1 - P_d)^2(1 - \eta)^{m_\abar + m_\bbar}[1 - (1 - P_d)(1 - \eta)^{m_a}][1 - (1 - P_d)(1 - \eta)^{m_b}]\\
  &= (1 - P_d)^2(1 - \eta)^{2x}[(1-\eta)^{-m_a} - (1 - P_d)][(1-\eta)^{-m_b} - (1-P_d)]
\end{split}
\end{equation}
using $m_a+m_\abar = m_b+m_\bbar =x$ to simplify. Unlike the 4 PNR situation (Case~1), we have not been able to derive a closed-form expression for Eq.~(\ref{eq:totalCoinc}) with the exact $c_2(\bm)$ above, due to the presence of the  exponentiated $m_a$ and $m_b$.
To linearize the equation, 
we make the approximation $(1-\eta)^{-m_i} \approx 1+m_i\eta$, valid for $m_i\eta \ll 1$ (and in turn requires $\mu\ll 1$).
We emphasize that this approximation is not desirable---$\eta= 1$ is of course the ambition of any photonic experiment---yet we nevertheless invoke it to reach an effective two-qubit density matrix, for the existence of a $\rho_i$ satisfying Eq.~(\ref{eq:rhoEff}) necessitates a coincidence formula \emph{linear} in the single-pair probabilities $\bp$. Incidentally, the fact such an approximation is necessary to derive a closed-form solution in Case 2 (as well as Cases 3 and 4 below)---but \emph{not} for Case 1---represents an interesting finding in its own right, and one of the key contributions our work.

Replacing $(1-\eta)^{-m_i} \approx 1+m_i\eta$ in Eq.~(\ref{eq:c2}), we thus arrive at
\begin{equation}
\label{eq:c2approx}
c_2(\bm) \approx (1 - P_d)^2(1 - \eta)^{2x}(P_d + m_a\eta)(P_d + m_b\eta).
\end{equation}
Summing this expression over $\bm$ [Eq.~(\ref{eq:multinomial})] and $x$ [Eq.~(\ref{eq:Poisson})] (see Appendix~\ref{app4Thr}), we find
\begin{equation}
\label{eq:C2}
C_2 \approx (1 - P_d)^2 e^{\mu[(1 - \eta)^2 - 1]}
         \Big\{\underbrace{[P_d + \mu\eta(1 - \eta)^2p_a][P_d + \mu\eta(1 - \eta)^2p_b]}_{\text{accidental coincidences}} + \underbrace{\mu\eta^2(1 - \eta)^2 p_{ab}}_{\text{correlated coincidences}}\Big\},
\end{equation}
whereby the same logic leading to Eqs.~(\ref{eq:rho1},\ref{eq:K1}) returns the effective density matrix
\begin{equation}
\label{eq:rho2}
\rho_2 = \frac{1}{K_2}\left\{[P_d\mathbbm{1}_A + \mu\eta(1 - \eta)^2\rho_A] \otimes [P_d\mathbbm{1}_B + \mu\eta(1 - \eta)^2\rho_B] + \mu\eta^2(1 - \eta)^2\rho_{AB} \right\},
\end{equation}
and 
\begin{equation}
\label{eq:K2}
K_2 = [2P_d + \mu\eta(1 - \eta)^2]^2 + \mu\eta^2(1 - \eta)^2.
\end{equation}

\subsection{Case 3 --- 2 PNR Detectors}
\label{sec:2PNR}
Here a single coincidence results when $\ket{a}$ and $\ket{b}$ record one click each. Since $\ket{\abar}$ and $\ket{\bbar}$ are not monitored, we sum over all $n_\abar$ and $n_\bbar$, yielding 
\begin{equation}
\begin{split}
\label{eq:c3}
c_3(\bm) & = \Pr(1|m_a) \left[\sum_{n_\abar=0}^\infty \Pr(n_\abar|m_\abar) \right] \Pr(1|m_b) \left[\sum_{n_\bbar=0}^\infty \Pr(n_\bbar|m_\bbar)\right] \\
 &= [P_d(1 - \eta)^{m_a} + (1 - P_d)m_a\eta(1 - \eta)^{m_a - 1}][P_d(1 - \eta)^{m_b} + (1 - P_d)m_b\eta(1 - \eta)^{m_b - 1}]\\
 &\approx [P_d(1 - m_a\eta) + (1 - P_d)m_a\eta][P_d(1 - m_b\eta) + (1 - P_d)m_b\eta],
\end{split}
\end{equation}
where $m_i\eta\ll 1$ is again taken to permit an analytical solution for the total coincidence probability, namely (Appendix~~\ref{app2PNR})
\begin{equation}
\label{eq:C3}
C_3 \approx \underbrace{[P_d + (1-2P_d)\mu\eta p_a][P_d + (1-2P_d)\mu\eta p_b]}_{\text{accidental coincidences}} + \underbrace{(1-2P_d)^2\mu\eta^2 p_{ab}}_{\text{correlated coincidences}},
\end{equation}
and hence the effective density matrix
\begin{equation}
\label{eq:rho3}
\rho_3 = \frac{1}{K_3} \left\{[P_d\mathbbm{1}_A + (1 - 2P_d)\mu\eta\rho_A] \otimes [P_d\mathbbm{1}_B + (1 - 2P_d)\mu\eta\rho_B] + (1 - 2P_d)^2\mu\eta^2\rho_{AB} \right\}
\end{equation}
with
\begin{equation}
\label{eq:K3}
K_3 = [2P_d + (1 - 2P_d)\mu\eta]^2 + (1 - 2P_d)^2\mu\eta^2.
\end{equation}

\subsection{Case 4 --- 2 Threshold Detectors} \label{sec:2thr}
In the fourth and final case of 2 threshold detectors, a single coincidence is logged when the detectors monitoring states $\ket{a}$ and $\ket{b}$ receive at least one click; as in Case~3, the absent detectors on the $\ket{\abar}$ and $\ket{\bbar}$ paths can be modeled by summing over all relevant outcomes. Therefore the coincidence probability conditioned on $\bm$ can be written as 
\begin{equation}
\begin{split}
\label{eq:c4}
c_4(\bm) &= \left[\sum_{n_a=1}^\infty \Pr(n_a|m_a) \right] \left[\sum_{n_\abar=0}^\infty \Pr(n_\abar|m_\abar) \right] \left[\sum_{n_b=1}^\infty \Pr(n_b|m_b) \right] \left[\sum_{n_\bbar=0}^\infty \Pr(n_\bbar|m_\bbar) \right] \\
&= [1 - (1 - P_d)(1 - \eta)^{m_a}][1 - (1 - P_d)(1 - \eta)^{m_b}]\\
&\approx [P_d + (1 - P_d)m_a\eta][P_d + (1 - P_d)m_b\eta],
\end{split}
\end{equation}
once again exploiting $m_i\eta\ll 1$. The corresponding total coincidence probability is then (Appendix~\ref{app2Thr})
\begin{equation}
\label{eq:C4}
C_4 \approx \underbrace{[P_d + (1 - P_d)\mu\eta p_a][P_d + (1 - P_d)\mu\eta p_b]}_{\text{accidental coincidences}} + \underbrace{(1 - P_d)^2\mu\eta^2 p_{ab}}_{\text{correlated coincidences}},
\end{equation}
while the density matrix is
\begin{equation}
\label{eq:rho4}
\rho_4 = \frac{1}{K_4}\left\{[P_d\mathbbm{1}_A + (1 - P_d)\mu\eta\rho_A] \otimes [P_d\mathbbm{1}_B + (1 - P_d)\mu\eta\rho_B] + (1 - P_d)^2\mu\eta^2\rho_{AB} \right\},
\end{equation}
with
\begin{equation}
\label{eq:K4}
K_4 = [2P_d + (1 - P_d)\mu\eta]^2 + (1 - P_d)^2\mu\eta^2.
\end{equation}
\end{widetext}

\section{Analysis} \label{sec:analysisresults}
\subsection{General Considerations}
As found in Sec.~\ref{sec:results}, Case~1 (4 PNR) remarkably admits an exact solution for the effective density matrix $\rho_1$ [Eqs.~(\ref{eq:rho1},\ref{eq:K1})], while the other three cases (4 threshold, 2 PNR, and 2 threshold) require assumptions on efficiency $\eta$ and generation rate $\mu$ in order to simplify to effective forms. Importantly, pushing the approximations even further such that $1-\eta\approx 1$, $1-P_d\approx 1$, and $1- 2P_d \approx 1$ reduces all four cases to 
\begin{equation}
\label{eq:reduced}
\rho_{\text{reduced}} \propto (P_d\mathbbm{1}_A + \mu\eta\rho_A) \otimes (P_d\mathbbm{1}_B + \mu\eta\rho_B)  + \mu\eta^2\rho_{AB},
\end{equation}
which is precisely the standard formula for noise in coincidence detection: the desired contribution scales like $\mu\eta^2$, with noise appearing as the product of the marginal states on the individual detectors~\cite{PhysRev.53.752, 10.1119/1.3354986}.

The extent to which each case deviates from this fully reduced approximation represents the main goal of the current paper. To assist in analyzing the validity and implications of the effective density matrices derived here, we now specialize to the single-pair case of a maximally entangled state, specifically $\rho_{AB} = \ket{\Phi^+}\bra{\Phi^+}$ with $\ket{\Phi^+} = \frac{1}{\sqrt{2}}(\ket{HH}+\ket{VV})$ and thus the marginal density matrices $\rho_A=\rho_B=\mathbbm{1}/2$. For such a state, we can define the visibility, fidelity, and concurrence for each case $i$ as
\begin{equation}
\label{eq:vis}
\cV_i = \frac{C_i(HH) - C_i(HV) - C_i(VH) + C_i(VV)}{C_i(HH) + C_i(HV) + C_i(VH) + C_i(VV)},
\end{equation}
\begin{equation}
\label{eq:fid}
\cF_i = \braket{\Phi^+|\rho_i|\Phi^+},
\end{equation}
\begin{equation}
\label{eq:con}
\mathfrak{C}_i = \max(0,\lambda_1-\lambda_2-\lambda_3-\lambda_4),
\end{equation}
respectively. Here we use the notation $C_i(ab)$ to describe the coincidence probability for a specific measurement setting $\ket{a}$ and $\ket{b}$, and $\{\lambda_1,\lambda_2,\lambda_3,\lambda_4\}$ denote the eigenvalues, in decreasing order, of the matrix $R_i=\sqrt{\sqrt{\rho_i} (\sigma_y\otimes\sigma_y) \rho_i^* (\sigma_y\otimes\sigma_y)\sqrt{\rho_i}}$~\cite{Wootters1998}. Because the coincidence probability can be calculated numerically via the full summation in Eq.~(\ref{eq:totalCoinc}), the visibility $\cV_i$ can be computed exactly; therefore we can leverage it to quantify the accuracy of the approximations leading to each effective density matrix $\rho_i$. Thereafter, we consider the density-matrix-specific metrics fidelity $\cF_i$ and concurrence $\mathfrak{C}_i$ to compare the relative performance of each detector configuration. For concreteness, we consider the nominal values $(P_d,\eta,\mu) = (10^{-6},0.1,0.02)$ as experimentally realistic conditions and vary one parameter at a time.

\subsection{Interferometric Visibility} \label{Section 4.1}
\begin{figure*}[!tb]
    \centering
    \includegraphics[width=0.7\textwidth]{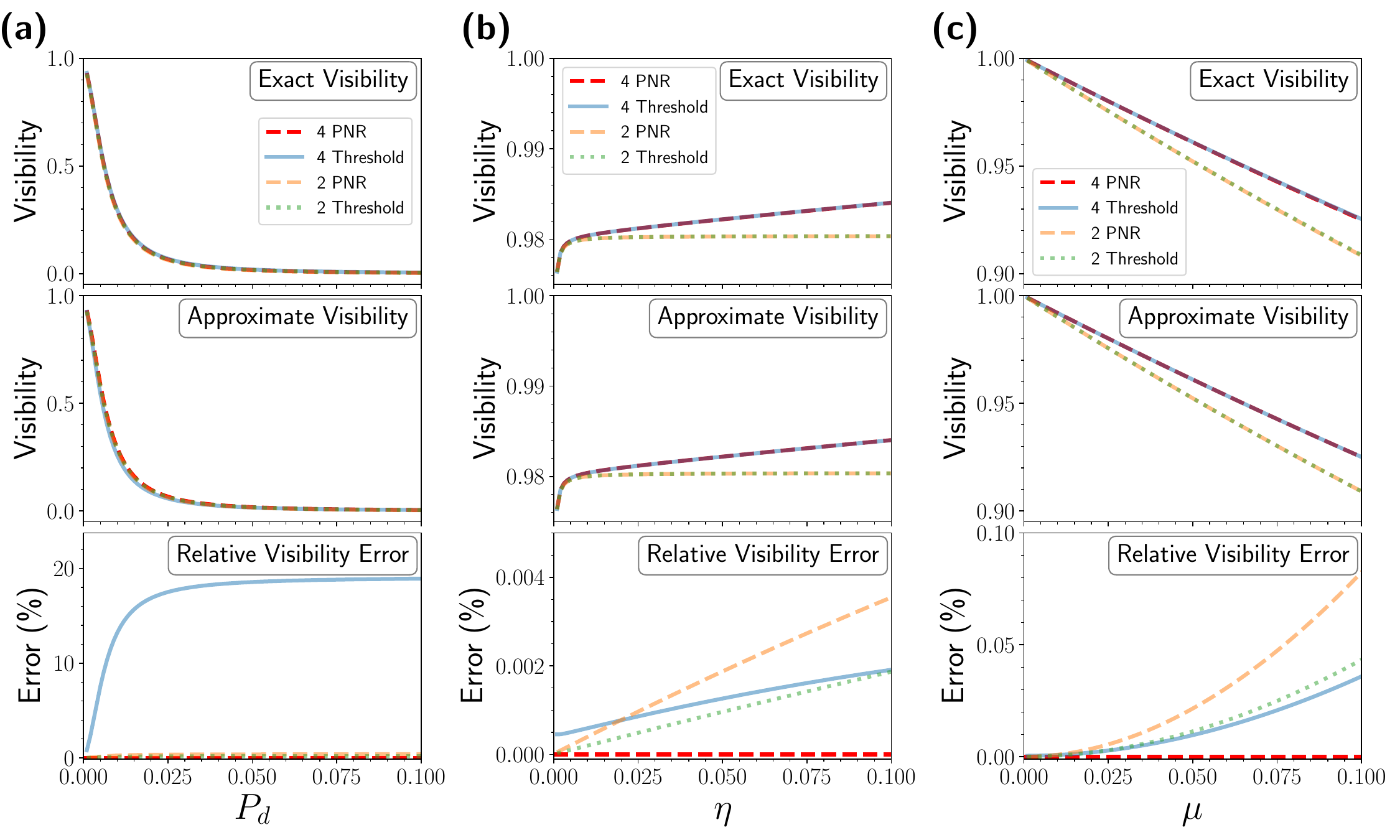}
    \caption{Comparison between exact and approximate visibilities from the model outlined in in Sec.~\ref{sec:results}. $(P_d,\eta,\mu) = (10^{-6}, 0.1,0.02)$ unless otherwise indicated on the $x$-axis: (a) the dark count probability $P_d$, (b) efficiency $\eta$, and (c) mean flux $\mu$.}
    \label{fig:fig2}
\end{figure*}

For the single pair state $\ket{\Phi^+}$, the coincidence probability $C_i(ab)$ can be computed for each setting in Eq.~(\ref{eq:vis}) by noting that $\bp=(0.5,0,0,0.5)$ for $\ket{ab}\in\{\ket{HH},\ket{VV}\}$ and $\bp=(0,0.5,0.5,0)$ for $\ket{ab}\in\{\ket{HV},\ket{VH}\}$; for an ideal case with no noise, this means $C_i(HV)=C_i(VH)=0$ and hence $\cV_i = 1$. To obtain the exact visibility, we evaluate Eq.~(\ref{eq:totalCoinc}) directly, truncated to a maximum of $x=10$ (which encompasses all probabilities $\Pr(x)$ for $\mu\leq0.1$ with less than $2.5\times 10^{-16}$ error).  
The approximate visibilities are computed using the probabilities in Eqs.~(\ref{eq:C1v1},\ref{eq:C2},\ref{eq:C3},\ref{eq:C4})---which for Case~1 is identical to the exact sum.

Figure~\ref{fig:fig2} plots the results for sweeping (a) $P_d$, (b) $\eta$, and (c) $\mu$ between 0 and 0.1 as three separate graphs each: the exact visibility (top), approximate visibility (middle), and the relative error (bottom). 
For the tests in Fig.~\ref{fig:fig2}(b,c), both the approximate and exact visibilities are high ($>$0.90) and in very good agreement (relative error $<$0.1\%), suggesting high accuracy for the approximations taken in the parameter range $\eta,\mu<0.1$.
In contrast, the error does approach 20\% for Case~2 in Fig.~\ref{fig:fig2}(a) (the $P_d$ scan). Such error does not suggest a poor model, though, but rather is an artifact of the low visibility. Both exact and approximate formulations predict $\cV_2<0.1$ for $P_d>0.02$; thus with relative error defined as $|\cV_2^{(\mathrm{approx})} - \cV_2^{(\mathrm{exact})}|/|\cV_2^{(\mathrm{exact})}|$, such low values of $\cV_2^{(\mathrm{exact})}$ amplify errors in a regime where the visibility is too low to be of practical utility.

Those interested in assessing model accuracy under different operating regimes are invited to test any set of parameters $(P_d,\mu,\eta)$ in the programs in our \href{https://github.com/ttruong1000/Developing-a-Practical-Model-for-Noise-in-Entangled-Photon-Detection}{GitHub repository} \cite{Truong2025}.

\subsection{Comparing Effective Density Matrices} \label{Section 4.3}
\begin{figure*}[!tb]
    \centering
    \includegraphics[width=0.7\textwidth]{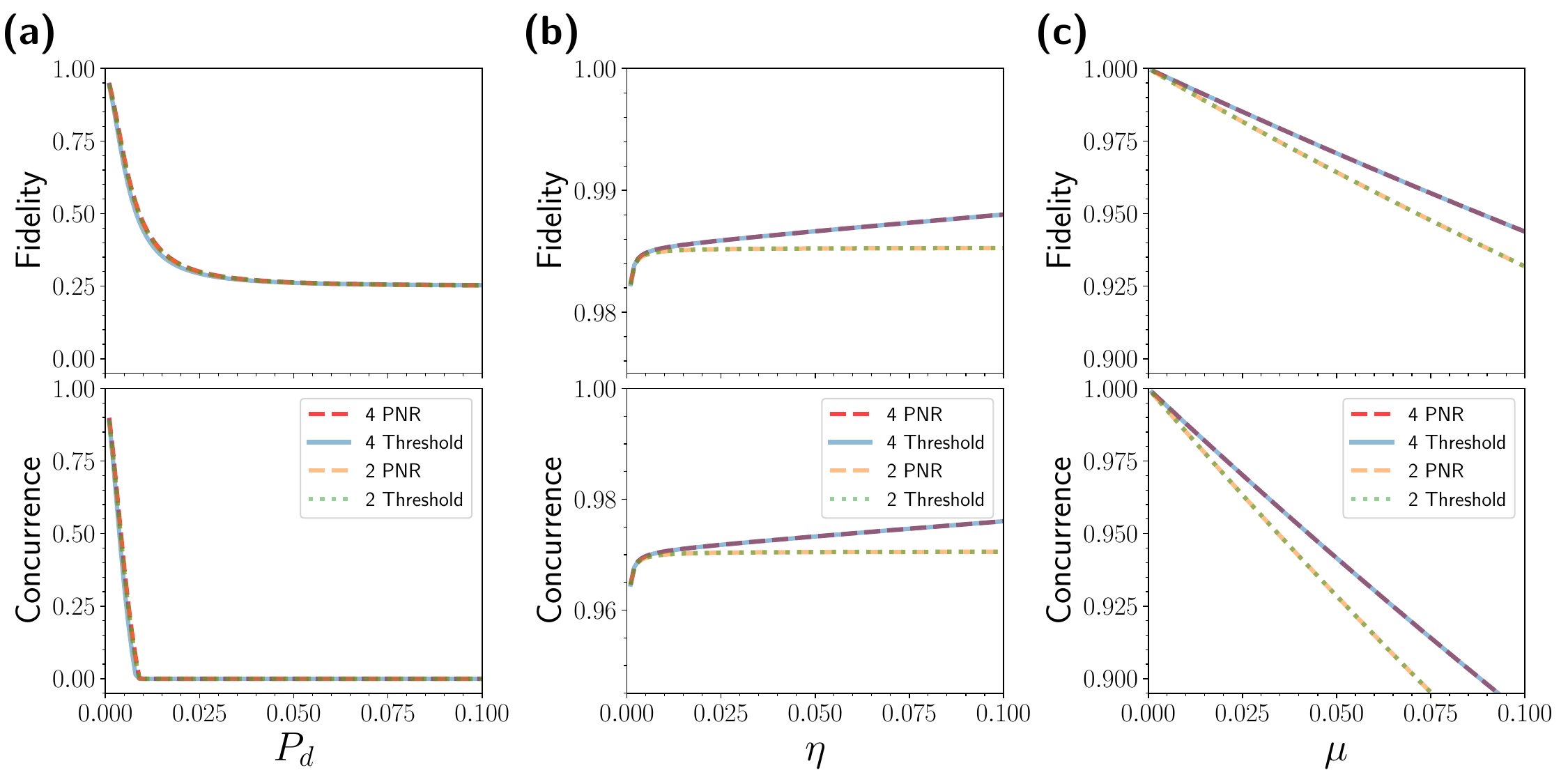}
    \caption{Fidelity and concurrence of the effective density matrices derived in Sec.~\ref{sec:results}. $(P_d,\eta,\mu) = (10^{-6}, 0.1, 0.02)$ except for the specific parameter tuned along the $x$ axis: (a) dark count probability $P_d$, (b) efficiency $\eta$, and (c) mean flux $\mu$.}
    \label{fig:fig4}
\end{figure*}

Much more interesting, however, is what is \emph{not} different in Fig.~\ref{fig:fig2}(b,c). Whereas the visibilities split between four- and two-detector cases as $\eta$ and $\mu$ increase, virtually no difference is seen between cases with the same number of PNR or threshold detectors. Intuitively, in the two-click coincidence experiment depicted in Fig.~\ref{fig:1}, the general motivation behind either adding detectors at $\ket{\abar}$ and $\ket{\bbar}$ or upgrading all detectors to PNR capabilities is to filter out spurious events in which coincidences at $\ket{a}$ and $\ket{b}$ do not correspond to photons from the same entangled pair. In the four-detector scenario, the registration of a click at either $\ket{\abar}$ or $\ket{\bbar}$ in tandem with clicks at $\ket{a}$ and $\ket{b}$ denotes either the detection of at least one dark count or the production of two photon pairs in the given timeslot; 
the simple strategy of throwing out any such event---which certainly may not prove optimal---leads to demonstrably higher visibilities for four detectors compared to two in the regimes of operation explored in Fig.~\ref{fig:fig2}(b,c).
On the other hand, under the same coincidence definition, PNR detectors show virtually no difference over the corresponding threshold configuration---a key finding of our study.

The equivalence between PNR and threshold detectors for the dark ports $\ket{\abar}$ and $\ket{\bbar}$ can be understood intuitively: since an $\ket{ab}$ coincidence requires these detectors to register vacuum, the capability to resolve higher-order photon events does not offer any benefit. Indeed, the ``no click'' probability for each class of detector is identical  under our model, namely $(1-P_d)(1-\eta)^{m_i}$. PNR detectors lead to coincidence probabilities different than threshold detectors only for events corresponding to two or more clicks on either $\ket{a}$ or $\ket{b}$, i.e., $\Pr(n_i|m_i)$ for $n_i\geq 2$, which under the approximations of interest for our matrix model (i.e., $\eta,\mu \ll 1$) are sufficiently rare to produce negligible differences in the visibilities recorded in Fig.~\ref{fig:fig2}. Of course, the situation can change markedly when either $\eta$ or $\mu$ is much larger, so our findings in no way diminish the overall value of PNR detectors in photonic quantum information processing. Yet it is interesting to find such negligible impact in the two-photon experiments of the form considered here.

The extremely low errors between the exact and approximate visibilities calculated in Sec.~\ref{Section 4.1}---$<$0.2\% for all cases except the high-$P_d$ settings of four threshold detectors as discussed---provide confidence in the approximations made to derive the effective density matrices in Eqs.~(\ref{eq:rho1}, \ref{eq:rho2}, \ref{eq:rho3}, \ref{eq:rho4}). Accordingly, we now shift to analyzing $\rho_i$ itself for each of the four cases $i\in\{1,2,3,4\}$, under parameter combinations $(P_d,\eta,\mu)$ validated in Fig.~\ref{fig:fig2}. Although the concurrence [Eq.~(\ref{eq:con})] requires numerical evaluation, fidelity $\cF_i = \braket{\Phi^+|\rho_i|\Phi^+}$ admits the closed-form expressions
\begin{align}
\cF_1 &= \frac{\left[P_d + \frac{1}{2}(1 - P_d)\mu\eta(1 - \eta)\right]^2 + (1 - P_d)^2\mu\eta^2}{\left[2P_d + (1 - P_d)\mu\eta(1 - \eta)\right]^2 + (1 - P_d)^2\mu\eta^2},\nonumber\\
\cF_2 &= \frac{\left[P_d + \frac{1}{2}\mu\eta(1 - \eta)^2\right]^2 + \mu\eta^2(1 - \eta)^2}{\left[2P_d + \mu\eta(1 - \eta)^2\right]^2 + \mu\eta^2(1 - \eta)^2}, \nonumber\\
\cF_3 & = \frac{\left[P_d + \frac{1}{2}(1 - 2P_d)\mu\eta\right]^2 + (1 - 2P_d)^2\mu\eta^2}{\left[2P_d + (1 - 2P_d)\mu\eta\right]^2 + (1 - 2P_d)^2\mu\eta^2}, \nonumber\\
\cF_4 &= \frac{\left[P_d + \frac{1}{2}(1 - P_d)\mu\eta\right]^2 + (1 - P_d)^2\mu\eta^2}{\left[2P_d + (1 - P_d)\mu\eta\right]^2 + (1 - P_d)^2\mu\eta^2}.
\end{align}

Figure~\ref{fig:fig4} plots fidelity and concurrence under the same settings explored in Sec.~\ref{Section 4.1}: nominally $(P_d,\eta,\mu)=(10^{-6},0.1,0.02)$, with single-parameter scans $P_d,\eta,\mu\in(0,0.1)$. The overall trends align fully with the visibility findings in Fig.~\ref{fig:fig2}, with a sharp drop in both $\cF_i$ and $\mathfrak{C}_i$ as $P_d$ increases and clear separation between four- and two-detector configurations in the $\eta$ and $\mu$ scans. Notably, $\mathfrak{C}_i=0$ for $P_d\gtrsim 0.01$ in all four cases, which validates the casual treatment of the high approximation error observed in the four-threshold case of Fig.~\ref{fig:fig2}(a), for it appears only in a regime where the entanglement vanishes and the state is of minimal practical value.

\begin{figure*}[!tb]
    \centering
    \includegraphics[width=0.7\textwidth]{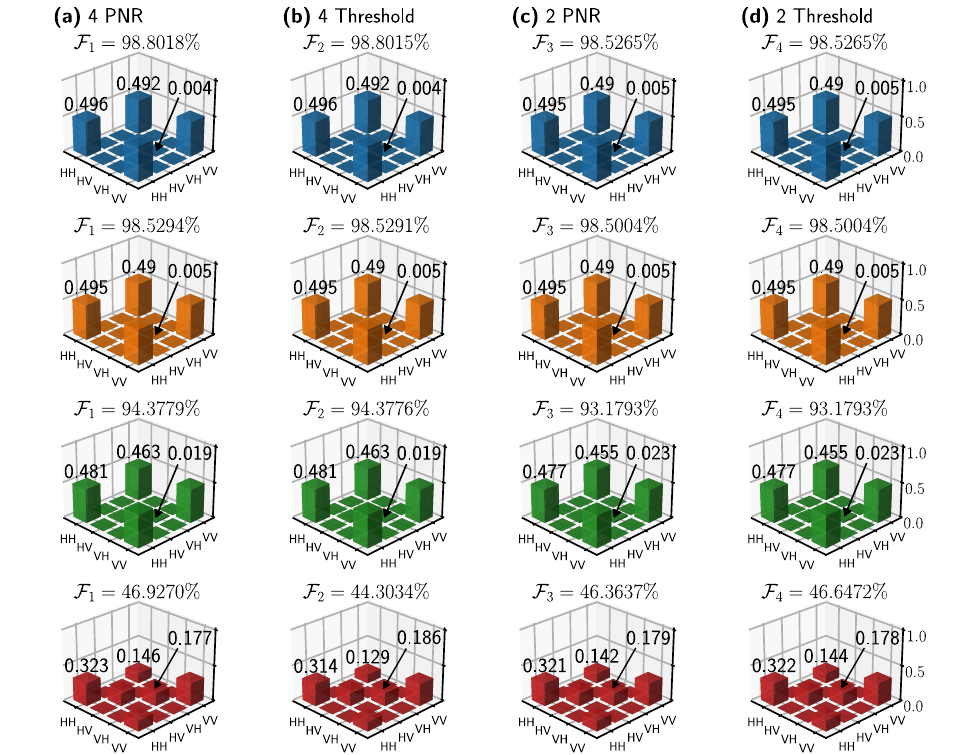}
    \caption{Effective density matrices $\rho_i$ for all four detector cases (left to right): (a) 4 PNR, (b) 4 threshold, (c) 2 PNR, and (d) 2 threshold. Parameter combinations $(P_d,\eta,\mu)$ from top to bottom in each (a--d) are $(10^{-6},0.1,0.02)$, $(10^{-6},0.01,0.02)$, $(10^{-6},0.1,0.1)$, and $(10^{-2},0.1,0.02)$. All imaginary components (not shown) are zero.}
    \label{fig:fig5}
\end{figure*}

For further insight into the effective density matrices, Fig.~\ref{fig:fig5} plots $\rho_i$ for four sets of parameters $(P_d,\eta,\mu)$: (a)~$(10^{-6},0.1,0.02)$, (b)~$(10^{-6},0.01,0.02)$, (c)~$(10^{-6},0.1,0.1)$, and (d)~$(10^{-2},0.1,0.02)$. Given the isotropic noise---due to identical detectors and equal probabilities for $\ket{H}$ and $\ket{V}$ in $\ket{\Phi^+}$---all states assume the standard Werner form $\rho_i = \lambda_i\ket{\Phi^+}\bra{\Phi^+} + \frac{1}{4}(1-\lambda_i)\mathbbm{1}_4$ differing only in mixing weight $\lambda_i$. Hence there exist only three unique nonzero values in each matrix, the values of which are annotated in Fig.~\ref{fig:fig5}: $\braket{HH|\rho_i|HH} = \braket{VV|\rho_i|VV}$, $\braket{HV|\rho_i|HV}=\braket{VH|\rho_i|VH}$, and $\braket{HH|\rho_i|VV} = \braket{VV|\rho_i|HH}$. In (a--c), the slight edge for four detectors over two appears in higher fidelity at either  the second (c) or third (a,b) significant digit; in (d), however, the four-threshold case possesses the lowest fidelity, likely another manifestation of the higher approximation error for this case at $P_d=0.01$.

\section{Conclusion} \label{sec:conclusion}
We have derived the total coincidence probabilities and effective density matrices for two-photon entanglement distribution under realistic experimental impairments---namely, probabilistic emission, nonunit efficiency, and dark counts. After proposing a general formalism applicable to either two or four PNR or threshold detectors, we specialize to independent Poisson-distributed photon pairs and obtain explicit formulas for the effective density matrices, under the condition of identical detectors for simplicity. The 4 PNR case admits exact results, whereas the other three configurations require approximations to the regime $\eta,\mu\ll 1$ in order to obtain closed-form solutions. Overall, we find four detectors offer noticeable improvements over two detectors in filtering out unwanted noise events, whereas PNR detectors reveal no significant advantages over threshold detectors in the studied regimes. Of course, these conclusions apply only to the case of two-photon detection where the odds of high-order contributions are intentionally kept low---historically the most common situation for SPDC. In contrast, for multiphoton experiments where higher photon numbers are desired, PNR detection is critical for measuring and exploiting the available optical resources. 

We can naturally extend the ideas presented here on multiple fronts. As noted in Sec.~\ref{sec:Prelim}, any dual-rail photonic encoding is included in the theory automatically. Similarly, high-dimensional $d$-rail qudit encodings can be handled by expanding from two to $d$-outcome measurements, increasing the length of the vectors $\bm$ and $\bn$ from four to $d^2$ but otherwise introducing no changes to the formalism. 
Additional adaptations are possible by modifying the form of the three PMFs in the fundamental model:
the channel efficiency and detector noise $\Pr(\bn|\bm)$ [Eq.~(\ref{eq:marginals})], the interpair correlations $\Pr(\bm|x)$ [Eq.~(\ref{eq:multinomial})], and the pair generation probability $\Pr(x)$ [Eq.~(\ref{eq:Poisson})].
As sketched by Eq.~(\ref{eq:darkCount}), integration of generic dark count models into $\Pr(\bn|\bm)$ is formally straightforward, and modifying the pair generation probability $\Pr(x)$ poses no major difficulties in the procedure.

On the other hand, alternative models for the ground truth photon distribution $\Pr(\bm|x)$ can add significant mathematical complexities. In particular, the photon-pair independence assumed in the full density matrix $\rho_{\mathrm{full}}$ [Eq.~(\ref{eq:tensorProd})] leads to a convenient closed-form expression [Eq.~(\ref{eq:multinomial})] that depends only on individual pair probabilities and retains the same form for any qubit measurement setting $\{\ket{a},\ket{\abar}\}$ and $\{\ket{b},\ket{\bbar}\}$. By contrast, photons that populate the same spectro-temporal modes lead to probability distributions $\Pr(\bm|x)$ that vary strongly with interference effects.

Although our combinatorial approach can encompass these situations with further extensions modeled after, e.g., the indistinguishable pair case in Ref.~\cite{Takesue2009EffectsOM}, such complex scenarios
can perhaps be most efficiently modeled through the formalism of GBS, i.e., a bank of squeezed-state inputs that are acted on by a linear circuit and PNR detection~\cite{Hamilton2017, Kruse2019}. 
Applying GBS mathematical tools~\cite{Quesada2018, Gagatsos2019, Quesada2019,  Su2019, Pizzimenti2021, Quesada2022} could therefore prove quite useful in further extensions of our effective density matrix approach. Such a generalized model could, perhaps, answer the question whether effective density matrices \`{a} la Eq.~(\ref{eq:rhoEff}) are even definable in the high-efficiency ($\eta\rightarrow 1$), high-flux ($\mu\gtrsim 1$) contexts to which GBS experiments aspire.

\begin{acknowledgments}
A portion of this work work was performed in part at Oak Ridge National Laboratory, which is managed by UT-Battelle LLC, for the U.S. Department of Energy under contract no. DE-AC05-00OR22725. Funding was provided by the U.S. Department of Energy, Office of Science, Advanced Scientific Computing Research (Field Work Proposal ERKJ432, Grant No. DE-SC0024257). Plots in this paper, as well as other plots that explore the shape of the graphs of each equation in Section 3, are available in the following \href{https://github.com/ttruong1000/Developing-a-Practical-Model-for-Noise-in-Entangled-Photon-Detection}{GitHub repository} \cite{Truong2025}.
\end{acknowledgments}

\section*{Data Availability}
The data that support the findings of this article are openly available \cite{Truong2025}. 

\appendix
\section{Mathematical Details} \label{Appendix}
In this extra section, we detail the algebraic manipulations necessary to obtain the desired results in Sec.~\ref{sec:results}. For the multinomial distribution in Eq.~(\ref{eq:multinomial}), we first note several expectation values that will prove useful below:
\begin{align}
\label{eq:expectations}
\braket{m_{ab}} & = x p_{ab}, & \braket{m_{ab}m_{a\bbar}} & = xp_{ab}p_{a\bbar}(x-1), \nonumber\\
\braket{m_{a\bbar}} & = x p_{a\bbar}, & \braket{m_{ab}m_{\abar b}} & = xp_{ab}p_{\abar b}(x-1), \nonumber\\
\braket{m_{\abar b}} & = x p_{\abar b},  & \braket{m_{a\bbar}m_{\abar b}} & = xp_{a\bbar}p_{\abar b}(x-1), \nonumber \nonumber\\
\braket{m_{ab}^2} & = xp_{ab}+x p_{ab}^2(x-1). & &
\end{align}

\begin{widetext}
\subsection{Case 1 --- 4 PNR Detectors} \label{app4PNR}
We start by expanding Eq.~(\ref{eq:c1}) as
\begin{equation}
\begin{split}
\label{eq:c1app}
c_1(\bm) & = (1 - P_d)^2(1 - \eta)^{2(x - 1)} [P_d(1-\eta) + (1 - P_d)\eta m_a][P_d(1-\eta) + (1 - P_d)\eta m_b] \\
& = (1 - P_d)^2(1 - \eta)^{2(x - 1)} [P_d^2(1-\eta)^2 + P_d(1 - P_d)\eta(1-\eta)(m_a+m_b) + (1-P_d)^2\eta^2m_a m_b]\\
& = (1 - P_d)^2(1 - \eta)^{2(x - 1)} [P_d^2(1-\eta)^2 + P_d(1 - P_d)\eta(1-\eta)(2m_{ab}+m_{a\bbar} + m_{\abar b}) \\
& \qquad\qquad\qquad\qquad\qquad\qquad + (1-P_d)^2\eta^2(m_{ab}^2 + m_{ab}m_{a\bbar} + m_{ab}m_{\abar b} + m_{a\bbar}m_{\abar b})],
\end{split}
\end{equation}
where the last line makes use of Eq.~(\ref{eq:singlePhotons}). Summing over all possible $\bm$ for a fixed $x$ allows us to leverage Eq.~(\ref{eq:expectations}) such that
\begin{equation}
\begin{split}
\label{eq:c1sum}
\sum_{\bm(x)} c_1(\bm) \Pr(\bm|x) & = (1 - P_d)^2(1 - \eta)^{2(x - 1)} [P_d^2(1-\eta)^2 + P_d(1 - P_d)\eta(1-\eta)\braket{2m_{ab}+m_{a\bbar} + m_{\abar b}} \\[-1em]
& \qquad\qquad\qquad + (1-P_d)^2\eta^2\braket{m_{ab}^2 + m_{ab}m_{a\bbar} + m_{ab}m_{\abar b} + m_{a\bbar}m_{\abar b}}]\\
& = (1 - P_d)^2(1 - \eta)^{2(x - 1)} \{P_d^2(1-\eta)^2 + P_d(1 - P_d)\eta(1-\eta)x(2p_{ab}+p_{a\bbar} + p_{\abar b}) \\
& \qquad + (1-P_d)^2\eta^2x[p_{ab} + p_{ab}^2(x-1) + p_{ab}p_{a\bbar}(x-1) + p_{ab}p_{\abar b}(x-1) + p_{a\bbar}p_{\abar b}(x-1)]\}\\
& = (1 - P_d)^2(1 - \eta)^{2(x - 1)} \{P_d^2(1-\eta)^2 + P_d(1 - P_d)\eta(1-\eta)x(2p_{ab}+p_{a\bbar} + p_{\abar b}) \\
& \qquad + (1-P_d)^2\eta^2[x(p_{ab} - p_{ab}^2 - p_{ab}p_{a\bbar} - p_{ab}p_{\abar b} - p_{a\bbar}p_{\abar b}) + x^2(p_{ab}^2 + p_{ab}p_{a\bbar} + p_{ab}p_{\abar b} + p_{a\bbar}p_{\abar b})]\},
\end{split}
\end{equation}

The subsequent summation over $x$ [Eq.~(\ref{eq:totalCoinc})] is facilitated by the relation
\begin{equation}
\label{eq:PoissShuff}
(1-\eta)^{2x}\Pr(x) = (1-\eta)^{2x}e^{-\mu}\frac{\mu^x}{x!} = e^{\mu[(1-\eta)^2-1]}\left\{e^{-\mu(1-\eta)^2} \frac{[\mu(1-\eta)^2]^x}{x!}\right\},
\end{equation}
where the factor in braces corresponds to the PMF of a Poisson distribution with mean $\mu(1-\eta)^2$.
Consequently, we can read off the sum over $x$ directly by replacing $x$ with $\mu(1-\eta)^2$ and $x^2$ with $\mu(1-\eta)^2+\mu^2(1-\eta)^4$:
\begin{equation}
\begin{split}
\label{eq:C1app}
C_1 & = \left(\frac{1-P_d}{1-\eta}\right)^2 e^{\mu[(1-\eta)^2-1]} \Big[P_d^2(1-\eta)^2 
 + P_d(1 - P_d)\mu\eta(1-\eta)^3(2p_{ab}+p_{a\bbar} + p_{\abar b}) \\
& \qquad + (1-P_d)^2\eta^2\{\mu(1-\eta)^2(p_{ab} - p_{ab}^2 - p_{ab}p_{a\bbar} - p_{ab}p_{\abar b} - p_{a\bbar}p_{\abar b}) \\
& \qquad\qquad\qquad\qquad + [\mu(1-\eta)^2+\mu^2(1-\eta)^4](p_{ab}^2 + p_{ab}p_{a\bbar} + p_{ab}p_{\abar b} + p_{a\bbar}p_{\abar b})\}\Big]\\
& = (1-P_d)^2 e^{\mu[(1-\eta)^2-1]} \Big\{P_d^2 
 + P_d(1 - P_d)\mu\eta(1-\eta)(2p_{ab}+p_{a\bbar} + p_{\abar b}) \\
& \qquad + (1-P_d)^2\eta^2[\mu p_{ab} 
  + \mu^2(1-\eta)^2(p_{ab}^2 + p_{ab}p_{a\bbar} + p_{ab}p_{\abar b} + p_{a\bbar}p_{\abar b})]\Big\}\\
& = (1-P_d)^2 e^{\mu[(1-\eta)^2-1]} \Big\{P_d^2 
 + P_d(1 - P_d)\mu\eta(1-\eta)(p_a + p_b)  + (1-P_d)^2\eta^2[\mu p_{ab} 
  + \mu^2(1-\eta)^2p_a p_b]\Big\}\\
& = (1 - P_d)^2e^{\mu[(1 - \eta)^2 - 1]} 
\Big\{[P_d + (1 - P_d)\mu\eta(1 - \eta)p_a][P_d + (1 - P_d)\mu\eta(1 - \eta)p_b] + (1 - P_d)^2\mu\eta^2 p_{ab}\Big\},
\end{split}
\end{equation}
matching Eq.~(\ref{eq:C1v2}). Conveniently, by  substituting in $p_a=p_{ab}+p_{a\bbar}$ and $p_b=p_{ab} + p_{\abar b}$ [Eq.~(\ref{eq:marginalP})] for all terms except the one scaling like the desired $\mu p_{ab}$, the noise contribution can be converted into product of factors depending on $p_a$ or $p_b$ only, subsequently facilitating the effective density matrix in Eq.~(\ref{eq:rho1}).

\subsection{Case 2 --- 4 Threshold Detectors} \label{app4Thr}
Expanding Eq.~(\ref{eq:c2approx}), we find
\begin{equation}
\begin{split}
\label{eq:c2appapprox}
c_2(\bm) & \approx (1 - P_d)^2 (1 - \eta)^{2x} (P_d + m_a\eta)(P_d + m_b\eta) \\
&\approx (1 - P_d)^2 (1 - \eta)^{2x} [P_d^2 + P_d\eta(m_a + m_b) + \eta^2m_am_b] \\
&\approx (1 - P_d)^2 (1 - \eta)^{2x} [P_d^2 + P_d\eta(2m_{ab} + m_{a\bbar} + m_{\abar b}) + \eta^2(m_{ab}^2 + m_{ab}m_{a\bbar} + m_{ab}m_{\abar b} + m_{a\bbar}m_{\abar b})],
\end{split}
\end{equation}
again leveraging Eq.~(\ref{eq:singlePhotons}).
Summing over $\bm$ and again invoking Eq.~(\ref{eq:expectations}):
\begin{equation}
\begin{split}
\label{eq:c2approxsum}
\sum_{\bm(x)} c_2(\bm) \Pr(\bm|x) & \approx (1 - P_d)^2(1 - \eta)^{2x} [P_d^2 + P_d\eta\braket{2m_{ab}+m_{a\bbar} + m_{\abar b}} \\[-1em]
& \qquad + \eta^2\braket{m_{ab}^2 + m_{ab}m_{a\bbar} + m_{ab}m_{\abar b} + m_{a\bbar}m_{\abar b}}]\\
& \approx (1 - P_d)^2(1 - \eta)^{2x} \{P_d^2 + P_d\eta x(2p_{ab}+p_{a\bbar} + p_{\abar b}) \\
& \qquad + \eta^2x[p_{ab} + p_{ab}^2(x-1) + p_{ab}p_{a\bbar}(x-1) 
+ p_{ab}p_{\abar b}(x-1) + p_{a\bbar}p_{\abar b}(x-1)]\}\\
& \approx (1 - P_d)^2(1 - \eta)^{2x} \{P_d^2 + P_d\eta x(2p_{ab}+p_{a\bbar} + p_{\abar b}) \\
& \qquad + \eta^2[x(p_{ab} - p_{ab}^2 - p_{ab}p_{a\bbar} - p_{ab}p_{\abar b}  - p_{a\bbar}p_{\abar b}) 
+ x^2(p_{ab}^2 + p_{ab}p_{a\bbar} + p_{ab}p_{\abar b} + p_{a\bbar}p_{\abar b})]\}.
\end{split}
\end{equation}
By Eq.~(\ref{eq:PoissShuff}), we again complete the sum over $x$ by replacing $x$ with $\mu(1-\eta)^2$ and $x^2$ with $\mu(1-\eta)^2+\mu^2(1-\eta)^4$:
\begin{equation}
\begin{split}
\label{eq:C2app}
C_2 & \approx (1-P_d)^2 e^{\mu[(1-\eta)^2-1]} \Big[P_d^2\mu\eta(1-\eta)^2(2p_{ab}+p_{a\bbar} + p_{\abar b}) \\
& \qquad + \eta^2\{\mu(1-\eta)^2(p_{ab} - p_{ab}^2 - p_{ab}p_{a\bbar} - p_{ab}p_{\abar b} - p_{a\bbar}p_{\abar b}) \\
& \qquad\qquad + [\mu(1-\eta)^2+\mu^2(1-\eta)^4](p_{ab}^2 + p_{ab}p_{a\bbar} + p_{ab}p_{\abar b} + p_{a\bbar}p_{\abar b})\}\Big]\\
& \approx (1-P_d)^2 e^{\mu[(1-\eta)^2-1]} \Big\{P_d^2 
 + P_d\mu\eta(1-\eta)^2(2p_{ab}+p_{a\bbar} + p_{\abar b}) \\
& \qquad + \eta^2[\mu(1-\eta)^2 p_{ab} 
  + \mu^2(1-\eta)^4(p_{ab}^2 + p_{ab}p_{a\bbar} + p_{ab}p_{\abar b} + p_{a\bbar}p_{\abar b})]\Big\}\\
& \approx (1-P_d)^2 e^{\mu[(1-\eta)^2-1]} \Big\{P_d^2 
 + P_d\mu\eta(1-\eta)^2(p_a + p_b) + \eta^2[\mu(1-\eta)^2 p_{ab} 
  + \mu^2(1-\eta)^4p_a p_b]\Big\}\\
& \approx (1 - P_d)^2e^{\mu[(1 - \eta)^2 - 1]}\Big\{[P_d + \mu\eta(1 - \eta)^2p_a][P_d + \mu\eta(1 - \eta)^2p_b] + \mu\eta^2(1-\eta)^2 p_{ab}\Big\},
\end{split}
\end{equation}
matching Eq.~(\ref{eq:C2}). 

\subsection{Case 3 --- 2 PNR Detectors} \label{app2PNR}
Expanding Eq.~(\ref{eq:c3}),
\begin{equation}
\begin{split}
\label{eq:c3appapprox}
c_3(\bm) 
& \approx [P_d(1 - m_a\eta) + (1 - P_d) m_a\eta][P_d(1 - m_b\eta) + (1 - P_d) m_b\eta] \\
& \approx P_d^2 + P_d(1 - 2P_d)\eta (m_a + m_b) + (1 - 2P_d)^2\eta^2 m_am_b \\
& \approx P_d^2 + P_d(1 - 2P_d)\eta(2m_{ab} + m_{a\bbar} + m_{\abar b} )
+ (1 - 2P_d)^2\eta^2(m_{ab}^2 + m_{ab}m_{a\overline{b}} + m_{ab}m_{\abar b}  + m_{a\overline{b}}m_{\overline{a}b}),
\end{split}
\end{equation}
and summing over $\bm$, we find
\begin{equation}
\begin{split}
\label{eq:c3approxsum}
\sum_{\bm(x)} c_3(\bm) \Pr(\bm|x) & \approx P_d^2 + P_d(1 - 2P_d)\eta\braket{2m_{ab}  + m_{a\bbar} + m_{\abar b}} \\[-1em]
& \qquad \qquad + (1 - 2P_d)^2\eta^2\braket{m_{ab}^2  + m_{ab}m_{a\bbar} + m_{ab}m_{\abar b} + m_{a\overline{b}}m_{\overline{a}b}} \\
& \approx P_d^2 + P_d(1 - 2P_d)\eta x(2p_{ab}  + p_{a\bbar} + p_{\abar b}) \\
& \qquad \qquad + (1 - 2P_d)^2\eta^2x[p_{ab} + p_{ab}^2(x-1) + p_{ab}p_{a\bbar}(x-1) \\
& \qquad\qquad\;\;\;  + p_{ab}p_{\abar b}(x-1) + p_{a\bbar}p_{\abar b}(x-1)],\\
& \approx P_d^2 + P_d(1-2P_d)\eta x(2p_{ab} +p_{a\bbar} + p_{\abar b} ) \\
& \qquad + (1-2P_d)^2\eta^2[x(p_{ab} - p_{ab}^2 - p_{ab}p_{a\bbar} - p_{ab}p_{\abar b} - p_{a\bbar}p_{\abar b}) \\
& \qquad\qquad\qquad\qquad\;\;\; + x^2(p_{ab}^2 + p_{ab}p_{a\bbar} + p_{ab}p_{\abar b} + p_{a\bbar}p_{\abar b})].
\end{split}
\end{equation}
Unlike Cases 1 and 2 [Eqs.~(\ref{eq:c1sum},\ref{eq:c2approxsum})], $x$ no longer appears in an exponent, simplifying the $x$ sum to a simple expectation over $\Pr(x)$ in Eq.~(\ref{eq:Poisson}) such that $\braket{x}=\mu$ and $\braket{x}=\mu+\mu^2$:
\begin{equation}
\begin{split}
\label{eq:C3app}
C_3 & \approx P_d^2 + P_d(1-2P_d)\mu\eta(2p_{ab}+p_{a\bbar} + p_{\abar b}) \\
& \qquad + (1-2P_d)^2\eta^2[\mu(p_{ab} - p_{ab}^2  - p_{ab}p_{a\bbar} - p_{ab}p_{\abar b} - p_{a\bbar}p_{\abar b}) 
 + (\mu + \mu^2)(p_{ab}^2 + p_{ab}p_{a\bbar} + p_{ab}p_{\abar b} + p_{a\bbar}p_{\abar b})]\\
& \approx P_d^2 + P_d(1 - 2P_d)\mu\eta(2p_{ab} + p_{a\bbar} + p_{\abar b} ) 
 + (1 - 2P_d)^2\eta^2[\mu p_{ab} + \mu^2(p_{ab}^2 + p_{ab}p_{a\bbar}  + p_{ab}p_{\abar b}  + p_{a\overline{b}}p_{\overline{a}b})]\\
& \approx P_d^2 + P_d(1 - 2P_d)\mu\eta(p_a + p_b) + (1 - 2P_d)^2\eta^2(\mu p_{ab} + \mu^2p_ap_b)\\
& \approx [P_d + (1 - 2P_d)\mu\eta p_a][P_d + (1 - 2P_d)\mu\eta p_b] + (1 - 2P_d)^2\mu\eta^2p_{ab},
\end{split}
\end{equation}
matching Eq.~(\ref{eq:C3}). 
\subsection{Case 4 --- 2 Threshold Detectors} \label{app2Thr}
Starting with Eq.~(\ref{eq:c4}),
\begin{equation}
\begin{split}
\label{eq:c4appapprox}
c_4(\bm) 
& \approx [P_d + (1 - P_d)m_a\eta][P_d + (1 - P_d)m_b\eta] \\
& \approx P_d^2 + P_d(1 - P_d)\eta(m_a + m_b) + (1 - P_d)^2\eta^2m_am_b \\
& \approx P_d^2 + P_d(1 - P_d)\eta(2m_{ab} + m_{a\bbar} + m_{\abar b} )
+ (1 - P_d)^2\eta^2(m_{ab}^2 + m_{ab}m_{a\bbar} + m_{ab}m_{\abar b} + m_{a\bbar}m_{\abar b}),
\end{split}
\end{equation}
we sum over $\bm$,
\begin{equation}
\begin{split}
\label{eq:c4approxsum}
\sum_{\bm(x)} c_4(\bm) \Pr(\bm|x) & \approx P_d^2 + P_d(1 - P_d)\eta\braket{2m_{ab} + m_{a\overline{b}} + m_{\overline{a}b} } \\[-1em]
& \qquad \qquad + (1 - P_d)^2\eta^2\braket{m_{ab}^2  + m_{ab}m_{a\overline{b}} + m_{ab}m_{\overline{a}b} + m_{a\overline{b}}m_{\overline{a}b}} \\
& \approx P_d^2 + P_d(1 - P_d)\eta x(2p_{ab} + p_{a\overline{b}} + p_{\overline{a}b} ) \\
& \qquad \qquad + (1 - P_d)^2\eta^2x[p_{ab} + p_{ab}^2(x-1) + p_{ab}p_{a\bbar}(x-1) 
 + p_{ab}p_{\abar b}(x-1) + p_{a\bbar}p_{\abar b}(x-1)],\\
& \approx P_d^2 + P_d(1-P_d)\eta x(2p_{ab}+p_{a\bbar} + p_{\abar b}) \\
& \qquad + (1-P_d)^2\eta^2[x(p_{ab} - p_{ab}^2 - p_{ab}p_{a\bbar} - p_{ab}p_{\abar b}  - p_{a\bbar}p_{\abar b}) 
 + x^2(p_{ab}^2 + p_{ab}p_{a\bbar} + p_{ab}p_{\abar b} + p_{a\bbar}p_{\abar b})],
\end{split}
\end{equation}
and then over $x$ using Poisson expectation values to finally obtain Eq.~(\ref{eq:C4}) in the main text:
\begin{equation}
\begin{split}
\label{eq:C3app}
C_4 & \approx P_d^2 + P_d(1-P_d)\mu\eta (2p_{ab}+p_{a\bbar} + p_{\abar b}) \\
& \qquad + (1-P_d)^2\eta^2[\mu(p_{ab} - p_{ab}^2 - p_{ab}p_{a\bbar} - p_{ab}p_{\abar b} - p_{a\bbar}p_{\abar b}) 
 + (\mu + \mu^2)(p_{ab}^2 + p_{ab}p_{a\bbar} + p_{ab}p_{\abar b} + p_{a\bbar}p_{\abar b})]\\
& \approx P_d^2 + P_d(1 - P_d)\mu\eta(2p_{ab}  + p_{a\overline{b}} + p_{\overline{a}b}) 
 + (1 - P_d)^2\eta^2[\mu p_{ab} + \mu^2(p_{ab}^2+ p_{ab}p_{a\overline{b}}  + p_{ab}p_{\overline{a}b} + p_{a\overline{b}}p_{\overline{a}b})]\\
& \approx P_d^2 + P_d(1 - P_d)\mu\eta(p_a + p_b) + (1 - P_d)^2\eta^2(\mu p_{ab} + \mu^2p_ap_b)\\
& \approx [P_d + (1 - P_d)\mu\eta p_a][P_d + (1 - P_d)\mu\eta p_b] + (1 - P_d)^2\mu\eta^2p_{ab}.
\end{split}
\end{equation}
\end{widetext}

\bibliography{references}

\end{document}